\documentclass[twocolumn,aps,prb,showpacs]{revtex4}

\usepackage{placeins}
\usepackage{graphicx}
\usepackage{color}
\usepackage{amssymb}
\usepackage{amsmath}     

\newcommand{\bra}[1]{\left\langle#1\right|}
\newcommand{\ket}[1]{\left|#1\right\rangle}

\def\e{\varepsilon}

\def\G{{\cal G}\,}
\def\g{\gamma}

\begin{document}

\preprint{Version 26.04.2002}

\title{Quasistates in a ring coupled to a reservoir and their relation
  to the Dicke effect}

\author{Bernhard Wunsch}
\email{bwunsch@physnet.uni-hamburg.de}
\affiliation{I. Institute of Theoretical Physics, University
of Hamburg, Jungiusstr. 9, D--20355 Hamburg, Germany}

\author{Alexander Chudnovskiy}
\affiliation{I. Institute of Theoretical Physics, University
of Hamburg, Jungiusstr. 9, D--20355 Hamburg, Germany}

\date{\today}

\begin{abstract}
  \vspace{5mm} We study the energy spectrum and the persistent current
  in an ideal one-dimensional mesoscopic ring coupled to an external
  fermionic reservoir. The contact between ring and reservoir is
  described by a tunneling operator, which causes an indirect coupling
  between different ring states via states in the reservoir. For
  strong coupling to the reservoir new quasistates with sharp
  eigenenergies develop inside the ring.  The formation of long-living
  states at strong tunnel coupling to the reservoir is analogous to
  the Dicke effect in optics, that was recently investigated in
  context of resonant scattering and resonant tunneling in solid state
  systems.  Our model reproduces the results obtained in previous work
  based on the scattering matrix approach and furthermore it describes
  a new stable energy spectrum in the limit of strong coupling.
\end{abstract}


\maketitle

\section{Introduction}

Experiments on mesoscopic rings enable to study quantum effects based
on phase coherence.  Two of its proven manifestations are oscillations
in the conductance of open rings connected to leads
\cite{Washburn92:1311,Fuhrer01:822} and persistent currents inside
closed rings
\cite{Levy90:2074,Chandrasekhar91:3578,Mailly93:2020,Jariwala01:1594,Lorke00:2223},
both periodic with the magnetic field applied perpendicular to the
probe.  In some of the experiments both effects can be measured on the
same probe with the help of side gates that control the coupling
between ring and reservoirs \cite{Mailly93:2020,Fuhrer01:822}.  In
recent optical experiments the energy spectra of quantum rings were
studied \cite{Lorke00:2223,Warburton00:926}.

As phase coherence is the precondition of these phenomena, the
influence of decoherence is of major interest.  Recently, the
suppression of quantum coherence in a mesoscopic system due to its
coupling to an external macroscopic reservoir attracted much
attention.  If a small mesoscopic system (quantum dot, quantum ring)
is coupled by tunneling to an external reservoir of fermions (a lead)
a phenomenon of level attraction is known to occur, which results in
changes of occupation numbers, statistics of energy levels, and
eventually the transport properties through the mesoscopic device
\cite{Chudnovskiy01:165316,Chudnovskiy02:819}.

In the present paper we investigate the effects of level attraction
due to coupling to an external reservoir on the persistent current in
a mesoscopic ring.  We find that the tunnel coupling in general leads
to the suppression of the persistent current. However, with increasing
coupling, the effective level structure of the ring coupled to the
reservoir changes. Due to level mixing through the reservoir,
quasistates with sharp eigenenergies develop in the ring, which can be
related to the Dicke effect in optics \cite{Dicke53:472, Dicke54:99,
  Shahbazyan94:17123, Shahbazyan98:6642}.  Depending on the number of
ring states coupled to the reservoir this results in a nonzero
persistent current even at very large tunneling between the ring and
the reservoir.  The saturation value of the persistent current at
large tunneling is crucially affected by the detailed structure of the
tunneling matrix elements. A complete suppression of the persistent
current takes place only if all states of the ring are mixed by
tunneling.

\begin{figure}
\begin{center}
\includegraphics[angle=0,scale=.28]{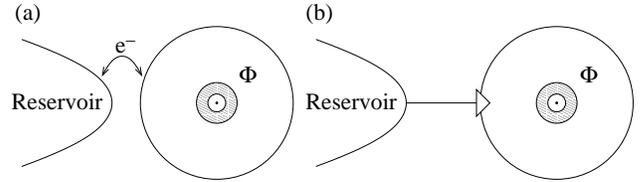}
\end{center}
\caption{Different setups for a coupled ring described within the tunnel Hamiltonian formalism (a)
  or within the scattering matrix approach (b).}
\label{fig:compButtikerTunnel}
\end{figure}

A ring coupled to a reservoir was investigated previously within the
scattering matrix approach \cite{Buttiker85:1846, Buttiker84:1982,
  Takai93:14318, Jayannavar94:13685, Akkermans91:76, Mello93:16358},
in which the ring is coupled via an ideal wire to the dissipative
reservoir see {Fig. \ref{fig:compButtikerTunnel}(b)}. But the
development of long living states for strong coupling was not
discussed there.

The work is organized as follows: After the introduction we explain
our model in the next section. In section III the general results for
the density of states (DOS) and the current density in the ring are
presented. Thereafter these results are analyzed for different numbers
of ring states that couple to the reservoir. In {section V} we relate
our results to the Dicke effect. A comparison with previous work based
on the scattering matrix approach will be performed in section VI.
Finally we will conclude our work.

\section{Model}

The setup studied in this work is shown in Fig.
\ref{fig:compButtikerTunnel} (a).  Since the main purpose of this
paper is to study the influence of decoherence introduced by a tunnel
contact to a fermionic reservoir (a lead) we confined ourselves to an
independent spinless electron model and assumed the ring to be
one-dimensional\cite{Cheung88:6050}.This simple model captures already
the main features of the energy spectrum and exhibits the persistent
current measured on rings in the ballistic transport
regime\cite{Mailly93:2020,Lorke00:2223}.  The Hamiltonian of our model
has the following form:
\begin{eqnarray}
\hat{H}&=& \sum_{m}\varepsilon_m \hat{a}_m^+
\hat{a}_m + \sum_{r} E_r \hat{b}_r^+ \hat{b}_r + \sum_{m,r}t_{m ,r} (\hat{a}_m^+ \hat{b}_r +
h. c.),\notag\\
\label{hamiltonian}
\end{eqnarray}
where $\hat{a}_m^+$ and $\hat{a}_m$ ($\hat{b}_r^+$ and $\hat{b}_r$)
are the creation and annihilation operators for electrons in the ring
(reservoir) with quantum number $m$ ($r$). The eigenfunctions of the
isolated ring are given by $\phi_m(\varphi)=e^{i\,m \varphi}$, where
$m$ denotes the angular momentum in the ring, and $\varphi$ is the
angular coordinate around the ring.  The corresponding eigenenergies
are given by $\varepsilon_m=4 E_0
\left(m+\frac{\Phi}{\Phi_0}\right)^2$, where $\Phi$ denotes the
magnetic flux through the ring, $\Phi_0=\frac{h}{e}$ is the magnetic
flux quantum, and the energy scale is given by $E_0=\frac{\hbar^2}{8
  m^* R^2}$. The flux dependence of the Hamiltonian as well as the
length of the ring is exclusively contained in the eigenenergies of
the ring.  The energies in the reservoir are denoted by $E_r$.

An important consequence of the coupling described by the tunneling
operator in {Eq. (\ref{hamiltonian})}, is that the angular momentum is
no longer conserved due to the new geometry, so that coupling to the
same states in the reservoir induces an indirect interaction between
the ring states.  It is this interaction that determines the behavior
of the system in the strong coupling regime.

Let us discuss briefly the effects of the neglected terms of the
Hamiltonian given by Eq. (\ref{hamiltonian}).  In the single particle
picture a ring of finite width can be solved analytically
\cite{Tan96:1635,Tan99:5626}. We have used the one-dimensional energy
spectrum to keep the calculations tractable, which is a good
approximation for thin rings.  The influence of electron-electron
interaction on the persistent current and the excitation spectrum has
been shown to be negligible in an ideal narrow ring
\cite{Chakraborty94:8460}. This is in agreement with experiments on a
single ring within the ballistic transport regime \cite{Mailly93:2020}
and with spectroscopy of nanoscopic semiconducting rings
\cite{Lorke00:2223}, for which the results can be explained within a
single particle picture.  Furthermore, for a high charge density
inside the ring the Coulomb interaction is screened and does not
contribute significantly.  The effect of spin is easily implemented in
our model as long as spin is conserved during the tunneling process.
Like in the isolated case \cite{Loss91:13762} the system can then be
described by an independent sum of a spin up subsystem and a spin-down
subsystem.

Following earlier work \cite{Shahbazyan98:6642} we now estimate the
tunneling matrix elements $t_{m,r}$ defined by $t_{m,r} =\bra{\phi_m}
V \ket{\psi_r}$, where the potential $V$ defines the region of overlap
between the wavefunctions in the ring and the reservoir and $\psi_r$
denotes an eigenfunction in the reservoir. For a small contact the
wavefunction of the reservoir can be taken out of the integral
$t_{m,r} \approx \psi_r(x_0) \int \phi_m^* (\varphi) V d\varphi$, so
that the dependence of the matrix element on the reservoir quantum
number is contained in a separate factor, that is independent of the
angular momentum. For the calculation of the Green's function of
electrons in the ring the matrix elements appear in pairs like
$t_{m_1,r} t_{m_2,r}^*$, so that the phase factor due the quantum
number of the reservoir cancels out. Assuming furthermore that
$|\psi_r(x_0)|^2$ is constant\cite{Shahbazyan98:6642}, the tunneling
matrix element is independent of $r$: $t_{m,r}=t_m$.

The dependence of the tunneling matrix element on the ring quantum
number can be estimated by inserting the eigenfunctions in the ring:
$t_m \propto \int_{-\varphi_0}^{\varphi_0}
e^{i\,m\varphi}=\frac{2}{m}\sin{m \varphi_0}$, where $\varphi_0$
describes the angular size of the contact.  For small $\varphi_0$ and
$m$ the coupling is independent of $m$, whereas it is suppressed for
higher $m$. In this paper we set the tunneling matrix elements
constant for a given range of angular momenta of the eigenstates
inside the ring. The tunneling matrix elements for other ring states
are set to zero.

\section{Methods and results}

Within the described model the DOS in the ring can be calculated for
arbitrary tunneling strength by means of a Dyson equation for the
Green's function. Therefore, the obtained results are also valid for
the strong coupling regime, in which the energy scale given by the
tunneling is of the order of or larger than the interlevel spacing
between consecutive ring states.  To avoid superimposing effects on
the DOS in the ring due to the band structure of the reservoir we
choose a constant density of states in the reservoir $\nu(E)=\nu$.
Setting $\hbar=1$ the Green's function for an electron in the ring has
the following form: {\small
\begin{eqnarray}
   \G_m(i E_n)=\G_m^0(i E_n)+
   \frac{\left[\G_m^0(i E_n)\right]^2 |t|^2 \g(i E_n)}
   {1- |t|^2 \g(i E_n)\sum_{m_1} (\G_{m_1}^0(i E_n))}
\end{eqnarray}
}
with
{\small
\begin{eqnarray}
   \g(i E_n)&=&\sum_r G_r^0(i E_n)=-i \pi \nu \,sign(E_n) \label{gamma},
\end{eqnarray}
} where $\G_m^0(i E_n)$ ($G_r^0(i E_n)$) denotes the Green's function
of the isolated ring state (isolated reservoir state) given by
$\G_m^0(i E_n)=(i E_n - \epsilon_m)^{-1}$ ($G_r^0(i E_n)= (i E_n -
E_r)^{-1}$), and $E_n$ denotes a Matsubara frequency. A different
density of states in the reservoir can be taken into account rather
straightforwardly as only the parameter $\gamma$ is changed in {Eq.
  (\ref{gamma})}.

Analytical continuation of the Green's function on the real axes leads
to the retarded Green's function, whose imaginary part determines the (normalized) spectral
density of the ring states $S_m(E)$ and the DOS in the ring $\rho(E)$
\begin{eqnarray}
     S_m(E)&=&-\frac{1}{\pi} \Im\left(\G^{ret}_m(E)\right)=\frac{\kappa}{\pi}\frac{\frac{1}{
     \left(E-\e_m\right)^2}}
{\left(1+\kappa^2 \xi(E)^2 \right)},\\
     \rho(E)&=&\sum_m S_m(E)=\frac{-\kappa \frac{\partial \xi(E)}{\partial E}}
     {\pi \left(1+\kappa^2 \xi(E)^2\right)}, \label{rhogeneral}
\end{eqnarray}
with
\begin{eqnarray}
  \xi(E)=\sum_{m_1}\frac{1}{E-\e_{m_1}}; \quad \kappa=\pi \nu |t|^2 \label{kappa}.
\end{eqnarray}

In the simplest case of only one ring state that couples to the
reservoir, the DOS is given by a Lorentzian centered around the
eigenenergy $\e_0$ of the isolated state $\rho(E)=\frac{\kappa}{\pi
  \left((E-\e_0)^2+ \kappa^2\right)}$ .  For any coupling strength,
the width of the maximum of the DOS is given by the coupling energy
$\kappa$.  For more coupling states and for strong coupling the
spectral densities are not given by Lorentzians due to the effective
interaction between different ring states, as is shown below.

The effect of coupling on the persistent current in the ring is
investigated by calculating the current density. It is obtained by
summing over the contributions of all ring states.  As the current
carried by an occupied isolated ring state is $I_m=-\frac{\partial
  \varepsilon_m}{\partial \Phi}$, the current density has the form:
\begin{eqnarray}
     j(E)&=&\sum_m -\frac{\partial \varepsilon_m}{\partial \Phi}S_m(E)=- \frac{\kappa \,\frac{\partial \xi(E)}{\partial \Phi}}
     {\pi(1+\kappa^2 \xi(E)^2)}\quad. \label{jgeneral}
\end{eqnarray}
The total
persistent current is then given by $I=\int_{-\infty}^{\infty} f(E) j(E)\,dE$,
where $f(E)=\frac{1}{1+exp\left(\beta(E-\mu)\right)}$ denotes the Fermi distribution.

An important consequence of a constant coupling strength for all
coupled ring states is that the ratio between current density and DOS
is independent of the coupling
\begin{eqnarray}
\frac{j(E)}{\rho(E)}=\frac{\frac{\partial \xi(E)}{\partial \Phi}}{\frac{\partial
\xi(E)}{\partial E}}.
\end{eqnarray}

\section{Analysis of the results for different numbers of coupled ring states}
In the following we will analyze the results obtained above for
different numbers of ring states that couple to the reservoir.

\subsection{Coupling of two ring states to the reservoir}
Now we assume that the coupling between ring and reservoir is
restricted to the two energetically lowest ring states.  This can be
motivated by selective tunneling with respect to the angular momentum
of the ring states as discussed above. Furthermore for a magnetic flux
close to $\Phi=\frac{\Phi_0}{2}$ this assumption is also a good
approximation as long as the energy gap to the higher lying ring
states is larger than the coupling energy $\kappa$. However we will
not limit the coupling strength in the following discussion.

\begin{figure}
\begin{center}
\includegraphics[angle=0,scale=0.9]{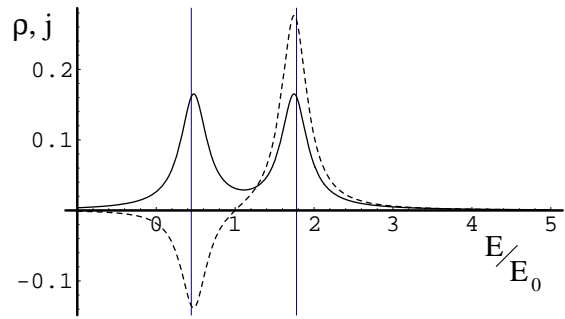}
\end{center}
\caption{DOS (full line) and current density (dashed line) for weak coupling
  $\kappa=0.2\,E_0$ and fixed magnetic flux $\Phi=\frac{\Phi_0}{3}$.
  The quasistates are energetically well separated and their spectral
  densities broaden with increasing coupling. The structures of
  current density and DOS are similar, which reflects low mixing of
  the ring states.  The grid lines show the eigenenergies of the
  isolated ring states.}
\label{fig:2N_klein}
\end{figure}

If only two ring states couple to the reservoir, the system can be
well described by introducing two quasistates.  For weak coupling,
their DOS is given by:
\begin{eqnarray}
S_{1,2}(E)&=&\frac{\kappa}{\pi\left[\left(E_{av} \pm \sqrt{\frac{(\Delta\e)^2}{4}-\kappa^2}\right)^2+\kappa^2\right]}
\end{eqnarray}
with
\begin{eqnarray}
\Delta\e=\e_{m_1}-\e_{m_2};\quad E_{av}=E-\frac{\e_{m_1}+\e_{m_2}}{2}.
\end{eqnarray}
The strength of coupling is characterized by the tunneling energy
$\kappa$ defined in {Eq. (\ref{kappa})}, which has to be compared with
the interlevel spacing $\Delta\e$ of the coupled ring states.  Without
coupling the quasistates coincide with the eigenstates of the isolated
ring.  {Fig. \ref{fig:2N_klein}} shows the DOS and the current density
in the weak coupling regime defined by
$\frac{(\Delta\e)^2}{4}-\kappa^2>0$. In this regime the spectral
densities of the quasistates broaden with increasing coupling, thereby
approaching each other.

\begin{figure}
\begin{center}
\includegraphics[angle=0,scale=0.9]{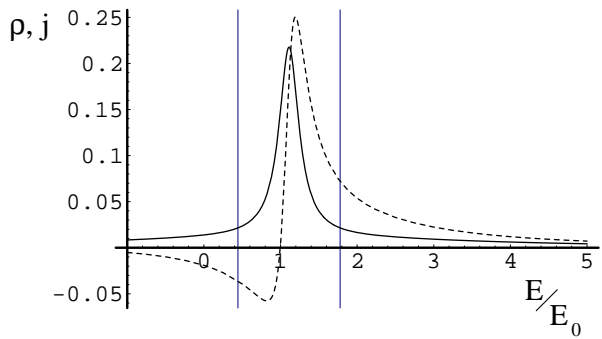}
\end{center}
\caption{ DOS (full line) and current density (dashed line) for
  strong coupling $\kappa=1.5\,E_0$ and fixed magnetic flux
  $\Phi=\frac{\Phi_0}{3}$. The DOS consists of a quasistate with a
  sharp eigenenergy and of a quasistate that contributes in a wide
  energy range. The asymmetric form of the current density (dotted
  line) differs substantially from the DOS, which shows strong mixing
  of the ring states.}
\label{fig:2N_gross}
\end{figure}

\begin{figure}
\begin{center}
  \includegraphics[angle=0,scale=0.9]{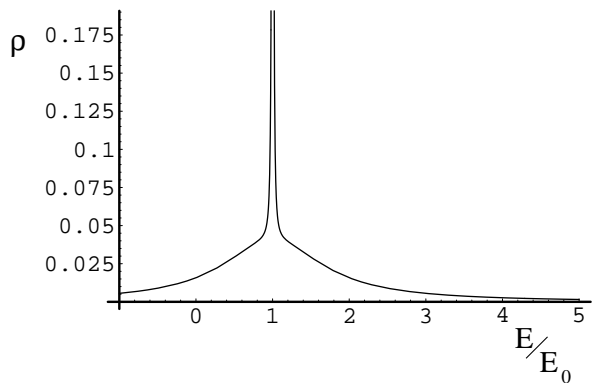}
\end{center}
\caption{DOS for strong coupling $\kappa=0.4\,E_0$ and fixed
  magnetic flux $\Phi=0.49\,\Phi_0$. For nearly degenerate
  eigenenergies of the isolated ring states the strong coupling regime
  is reached already for small coupling strength and the different
  behavior of the two quasistates is well resolved.}
\label{fig:2N_gross_dicke}
\end{figure}

At the critical coupling $\kappa_{c}=|\frac{\Delta\e}{2}|$ the
spectral densities of the quasistates are equal. Fig.
\ref{fig:2N_gross} and \ref{fig:2N_gross_dicke} illustrate the DOS and
the current density, if the coupling is increased to the strong
coupling regime with $\kappa>\kappa_c$. A new quasistate with a sharp
eigenenergy develops and is represented by the sharp peak in the DOS
with a width smaller than $\kappa$. The other quasistate contributes
to the DOS within a broad energy range of a width larger than
$\kappa$.

This behavior of the DOS is well described by the spectral densities
of the quasistates in the strong coupling regime: {\small
\begin{eqnarray}
S_{1,2}(E)=\frac{\kappa\mp\sqrt{\kappa^2-\frac{(\Delta\e)^2}{4}}}
{\pi\left[E_{av}^2+\left(\kappa \mp \sqrt{\kappa^2-\frac{(\Delta\e)^2}{4}} \right)^2\right]}
\quad.
\end{eqnarray}
}

It is important to notice that the energies of the isolated ring
states depend on the magnetic flux, while the coupling to the
reservoir is assumed to be independent of the magnetic flux. In
particular, the eigenenergies are degenerate at $\Phi=n
\frac{\Phi_0}{2}$, so that by changing the magnetic flux close to this
degenerate value one finally enters the regime of strong coupling, for
any nonzero coupling strength.
  
The energy of the long living state depends on the magnetic flux, as
it is given by the average energy of the two coupled states.
Therefore, the system shows Aharonov-Bohm type behavior even in the
strong coupling regime.  Correspondingly the persistent current
saturates in the limit of strong coupling and does not vanish.

The value of the saturated persistent current is obtained by noting
that the ratio between current density and DOS is independent of the
coupling.  In the limit of strong coupling, the long living state is
centered at the average energy of the coupling states and carries the
current $I=\frac{I_{m_1}+I_{m_2}}{2}$, while the strongly coupling
state carries a current of $I=\frac{I_{m_1}+I_{m_2}}{4}$.
 
\begin{figure}
\begin{center}
\includegraphics[angle=0,scale=0.9]{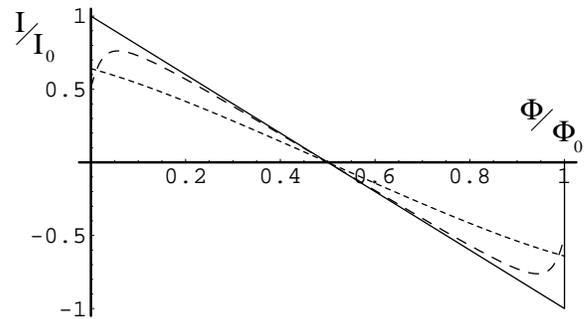}
\end{center}
\caption{Dependence of the persistent current on the magnetic flux through the ring 
  for a coupling of two ring states to the reservoir ($\mu=4\,E_0$).
  Relative to the persistent current in an isolated ring (full line),
  the persistent current is reduced and smoothed for increasing
  coupling (long dashed line $\kappa=0.2\,E_0$). In the strong
  coupling regime (short dashed line, $\kappa=5\,E_0$) the current
  saturates and the sawtooth form is restored.}
\label{fig:2N_I}
\end{figure}

Therefore, the value of the saturated persistent current is either
$\frac{1}{4}$ or $\frac{3}{4}$ of the current in the isolated ring,
depending on whether the Fermi energy lies below or above the energy
of the long-living state. For the latter case, \mbox{Fig.
  \ref{fig:2N_I}} shows the persistent current as a function of the
magnetic flux for different coupling parameters. In the strong
coupling regime (short dashed line) the current saturates and the
sawtooth form of the current is restored.

\subsection{Coupling of a finite number of ring states to the reservoir}

A generalization of the simplified two level model is obtained by
considering the coupling of more ring states to the reservoir. Thereby
at least all states with an eigenenergy below the Fermi energy are
coupled to the reservoir.

\begin{figure}
\begin{center}
\includegraphics[angle=0,scale=.9]{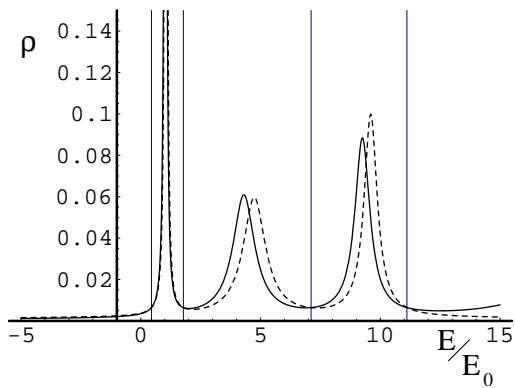}
\end{center}
\caption{DOS in the strong coupling regime $\kappa=5\,E_0$ for four coupled ring
  states (dotted line) and eight coupled states (full line). The
  smaller the interlevel spacing the more pronounced are the new
  quasistates. Magnetic flux $\Phi=\frac{\Phi_0}{3}$ is fixed.}
\label{fig:vNrho}
\end{figure}

The calculation of the DOS and the current density is performed in the
appendix.  In the strong coupling regime the system develops long
living states between the energetically adjacent states of the
isolated ring whenever the tunneling energy $\kappa$ exceeds the
interlevel spacing between the corresponding eigenenergies of the
isolated ring.  These new quasistates are more pronounced at small
energies as illustrated in {Fig. \ref{fig:vNrho}}. The reason is that
the energy scale connected with the coupling is the same for all
states $\kappa=\pi \nu |t|^2$ whereas the interlevel spacing between
adjacent ring states increases with their respective energies.

\begin{figure}
\begin{center}
\includegraphics[angle=0,scale=0.9]{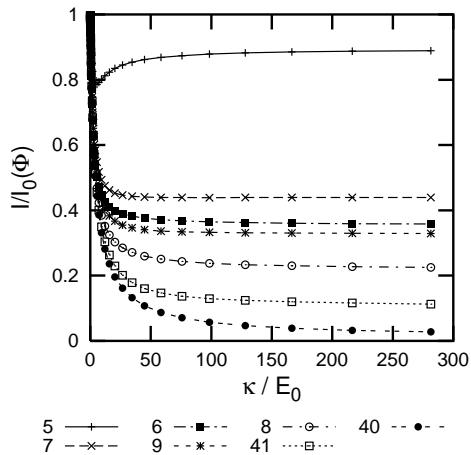}
\end{center}
\caption{Dependence of the persistent current on the coupling strength 
  for different numbers of coupled states and fixed magnetic flux
  $\Phi=\frac{\Phi_0}{3}$. The numbers of coupled states are listed
  according to the value at which the persistent current saturates.
  There are five eigenenergies of the isolated ring below the Fermi
  energy $\mu=25\,E_0$.  }
\label{fig:vNIk5}
\end{figure}

Like in the two level system the persistent current saturates in the
limit of strong coupling at a generally nonzero value.  Thereby the
saturation value of the persistent current depends strongly on the
number of coupled states.  It decreases with increasing number of
coupled ring states, but it also shows an odd-even effect with the
number of coupled states as illustrated in {Fig. \ref{fig:vNIk5}}.
Both features have their origin in the alternating sign of the current
carried by consecutive ring states.

\subsection{Coupling of all ring states to the reservoir}

According to our discussion of the tunneling matrix elements in
section II a coupling of all ring states to the reservoir is realized
in the limit of a point contact.  It is an appealing feature of our
model that a simple analytical formula for the DOS in the ring and the
current density can be given for this limiting case:
\begin{eqnarray}
\rho(E)=\frac{\kappa}{\pi} 
\frac{\frac{1}{x}\sin x \left(\cos\tilde{\Phi}-\cos\,x \right)+1-\cos\,x \cos\tilde{\Phi}}
{\kappa^2 \left(\sin\,x \right)^2 + \frac{4 E_0^2 x^2}{\pi^4}\left(\cos\tilde{\Phi}-\cos\,x \right)^2}
\label{rho}
\end{eqnarray}
with $x=\pi \sqrt{\frac{E}{E_0}}$ and the dimensionless flux $\tilde{\Phi}=2\pi\frac{\Phi}{\Phi_0}$.

\begin{figure}
\begin{center}
\includegraphics[angle=0,scale=.9]{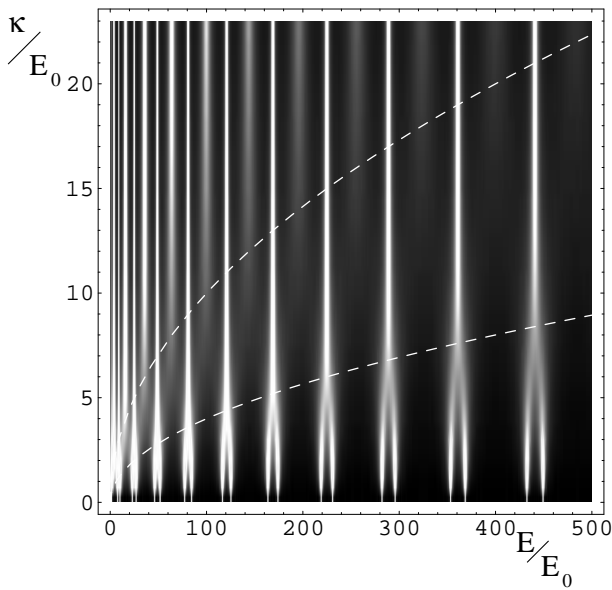}
\end{center}
\caption{Density plot of DOS as a function of the coupling for fixed
  magnetic flux $\Phi=0.4 \Phi_0$ (the magnitude of the DOS increases
  going from black to white).  For weak coupling the maxima are
  located at the eigenenergies of the isolated ring states whereas for
  strong coupling quasistates at $E=n^2\, E_0$ develop.  The critical
  coupling thereby depends on the energy as $\sqrt{E}$ with different
  prefactors for even or odd n as indicated by the dashed lines ($0.4
  \sqrt{E}$, $\sqrt{E}$).}
\label{fig:ANrhoEkappa}
\end{figure}

To obtain the DOS given in {Eq. (\ref{rho})} we have used
\begin{equation}
\xi(E) =
\sum_{m=-\infty}^{\infty} \frac{1}{E-\e_m} =\frac{\pi^2}{2 E_0 x}
\frac{\sin x}{\cos \tilde{\Phi} -\cos x}.
\end{equation}
Fig. \ref{fig:ANrhoEkappa} illustrates the development of the DOS with
increasing coupling. For small coupling the DOS shows Lorentz
broadened maxima around the eigenenergies of the isolated ring states
with a width given by the coupling energy $\kappa$. This indicates
that for small coupling each ring state couples independently to the
reservoir and does not interact with the other ring states. Analyzing
Eq. (\ref{rho}) the DOS shows maxima in the weak coupling regime at
the energies where the second term of the denominator disappears,
which happens exactly at the eigenenergies of the isolated ring
states. At the magnetic flux $\Phi=0.4\Phi_0$ used in the calculations
represented in {Fig.  \ref{fig:ANrhoEkappa}}, these eigenenergies are
grouped in pairs.  Each pair consists of ring states with angular
momenta $m$ and $-m-1$.

For strong coupling however the DOS forms sharp peaks at $E=n^2\,E_0$,
independently of the magnetic flux. These maxima are more and more
pronounced with increasing coupling. Analyzing again Eq. (\ref{rho})
the roots of the first term in the denominator determine the positions
of the maxima in the strong coupling regime.  The energies of the
quasistates can be understood from the following requirements.  The
quasistates lie energetically between each two neighbor states of the
isolated ring. Moreover, the suppression of persistent current and
Aharonov-Bohm effect demands the energies of the quasistates to be
independent of the magnetic flux. The positions of the quasistates
$E=n^2\,E_0$ are the only points satisfying both requirements above.
Therefore, no quasistates at other energies can form at strong
coupling.

The critical coupling at which the transition between weak and strong
coupling occurs depends on the interlevel spacing between the ring
states and hence on the energy and on the magnetic flux.  As
illustrated in \mbox{Fig. \ref{fig:ANrhoEkappa}} the critical coupling
grows with energy proportional to $\sqrt{E}$ and the quasistates are
developed first between pairs of ring states. For only two coupling
states it was shown that the critical coupling is given by
$\kappa_{c}=|\frac{\Delta\e}{2}|$, which is also a good estimation for
the critical coupling at which the quasistates at $E_{m}=(2m+1)^2 E_0$
develop:
\begin{eqnarray}
\kappa_c&\approx&\frac{\e_{-m-1}-\e_{m}}{2}=2(1-2\frac{\Phi}{\Phi_0})(2m+1)E_0\notag\\
&=&0.4 \sqrt{E_{m} E_0}\,.
\end{eqnarray}
As a guide to the eye, the dependence $\kappa_{c}(E_m)$ is shown by
the lower lying dashed line in Fig. \ref{fig:ANrhoEkappa}.
Correspondingly the second dashed line in Fig.  \ref{fig:ANrhoEkappa}
shows at which coupling strength the other half of long living states
develop at $E=(2n)^2\,E_0$.

\begin{figure}
\begin{center}
\includegraphics[angle=0,scale=0.7]{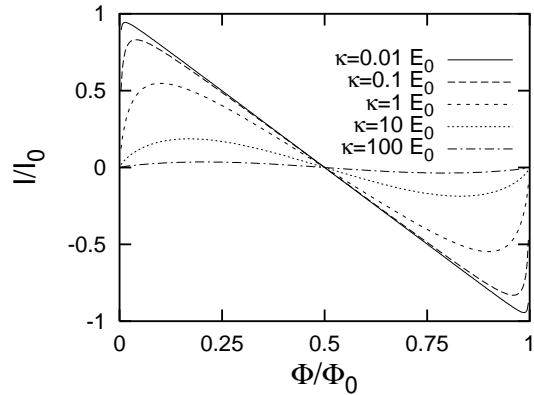}
\end{center}
\caption{Dependence of the persistent current on the magnetic flux 
  through the ring for different coupling strengths and for coupling
  of all ring states. The persistent current is continuously
  suppressed with increasing coupling and vanishes in the limit of
  strong coupling. For all lines, there are five states below the
  Fermi energy $\mu=25 E_0$.}
\label{fig:ANIphi_article}
\end{figure}

In contrast to the coupling of two ring states, the eigenenergies of
the long living states are now independent of the magnetic field,
which indicates the localization of those states.  Consequently, the
Aharonov-Bohm effect disappears, which is accompanied by a continuous
suppression of the persistent current with increasing coupling as
illustrated in Fig. \ref{fig:ANIphi_article}.

The current density inside the ring can be calculated with the help of Eq.  (\ref{jgeneral}) and is given by:
\begin{eqnarray}
j(E)&=& -\frac{\kappa}{\pi}\frac{\frac{\partial \xi(E)}{\partial \Phi}}{1+\kappa^2 \xi(E)^2} =\notag\\
&=&\frac{-4 \kappa \frac{E_0}{\pi^2 \Phi_0} x \sin x \sin{\tilde{\Phi}} } 
{\kappa^2 \left(\sin x \right)^2+\frac{4 E_0^2 x^2}{\pi^4} \left(\cos\tilde{\Phi} - \cos x \right)^2 }.
\label{ANj}
\end{eqnarray}

For weak coupling the current density shows Lorentz broadened maxima
with alternating sign around the eigenenergies of the isolated ring
states. In the strong coupling regime however it shows antisymmetric
peaks around the eigenenergies $E=n^2\,E_0$ of the newly evolved
quasistates.  Analyzing Eq. (\ref{ANj}) the current density has the
same denominator as the DOS given in \mbox{Eq.  (\ref{rho})}.
However, the numerator disappears at the eigenenergies $E=n^2 E_0$ of
the quasistates and therefore causes the antisymmetric peaks.

The asymmetric behavior of the current density causes a suppression of
the total persistent current with increasing coupling to the reservoir
which is illustrated in \mbox{Fig. \ref{fig:ANIphi_article}}.

The continuous suppression of the persistent current with increasing
coupling can also be understood with the help of the
coupling-independent ratio between current density and DOS, which is
given by
\begin{eqnarray}
\frac{j(E)}{\rho(E)}=\frac{-4 \frac{E_0}{\pi \Phi_0} x \sin{x} \sin{\tilde{\Phi}}}
{\frac{1}{x}\sin x \left(\cos\tilde{\Phi}-\cos x \right)+1-\cos x \cos\tilde{\Phi}}\quad.
\label{eq:alleNiveaus:ratio}
\end{eqnarray}
In contrast to the coupling of only two ring states this ratio
vanishes at eigenenergies of the quasistates that develop in the
strong coupling regime, so that eventually the persistent current will
also vanish in the limit of strong coupling. Therefore, the
quasistates do not carry current in contrast to the coupling of two
ring states where the long living quasistate carries the current
$I=\frac{I_{m_1}+I_{m_2}}{2}$.
 
\section{Connection to the Dicke effect}
Studying the energy spectrum of a ring coupled to a reservoir, we
showed that due to the coupling of different ring state coherent
collective states develop inside the ring. These quasistates lead to a
new sharp energy spectrum as illustrated in Fig.
\ref{fig:2N_gross_dicke}, \ref{fig:vNrho}, \ref{fig:ANrhoEkappa}.  The
same mechanism is known in optics as Dicke effect
\cite{Dicke53:472,Dicke54:99}.  Originally, the Dicke effect describes
how atoms in an atomic gas can form a coherent collective state when
they are coupled by an electromagnetic field with a wavelength bigger
than the distance between the radiating atoms. These collective states
have a sharp eigenenergy and lead to a significant reduction of the
Doppler width in atomic spectroscopy. In optics, the long living
states with a small spectral width are called subradiant and the broad
states that couple strongly to the light are called superradiant.

In our work different ring states are coupled by the fermionic
reservoir, which is therefore the counterpart of the electromagnetic
field in optics. 

Recently a similar system was investigated in the context of resonant
scattering \cite{Shahbazyan98:6642}. T. V. Shahbazyan and S. E. Ulloa
studied the electronic states of a system consisting of a 2D electron
gas, which is tunnel coupled to an array of pointlike, single level
quantum dots.  The localized states in the dots are coupled to each
other via the tunneling into and out of the delocalized states of the
2D gas.  Under certain conditions a coherent collective state is
formed that is located in the quantum dot array.

One can draw direct analogy between the physical entities entering the
model of Ref. \cite{Shahbazyan98:6642} and our model.  The states of
the locally distributed quantum dots correspond in our system to the
discrete ring states, whereas the 2D electron gas corresponds to the
fermionic reservoir.  In both models the underlying physics consists
in the formation of collective quasistates with narrow spectral
linewidths as the tunnel coupling exceeds some critical value.  In our
model the formation of the collective quasistates affects the
persistent current inside the loop, while the coherent collective
state located in the quantum dot array leads to a reduction of the
electron mobility in the 2D electron gas.

Furthermore, T. V. Shahbazyan and S. E. Ulloa showed that if the
discrete states in the quantum dots are energetically degenerate then
their DOS mainly consists of a sharp peak and a very low and broad
background. Thereby a fraction of up to $1-\frac{1}{N}$ of the states
contribute to the sharp peak in the DOS whereas the small remaining
part contributes to the DOS in a wide energy range
\cite{Shahbazyan98:6642}.  This is in agreement with our system, for
which in the strong coupling regime a single quasistate hybridizes
strongly with the reservoir and becomes extremely broad, whereas all
other quasistates show up as sharp maxima in the DOS.

However we want to stress two major differences to our work.  In our
work all ring states couple to the reservoir at the same point,
whereas in Ref. \cite{Shahbazyan98:6642} the interacting subsystems
are spatially separated. Therefore, the electron has to propagate
inside the 2D electron gas between consecutive tunneling events.  This
causes an additional phasefactor in the tunneling matrix elements that
destroys coherence. In order that coherent collective states are
formed, the average distance between different dots has to be of the
order of or smaller than the Fermi wavelength. Similarly, the original
Dicke effect in optics takes place only if the distance between the
atoms of the gas is of the order of or smaller than the wavelength of
the light.  This additional phase coherence length is absent in our
model, as all the ring states are localized within the same small
volume.

Furthermore, within our model the interlevel spacing between the ring
states can be adjusted systematically by changing the magnetic field.
Consequently, the regime of strong coupling can be reached by changing
the magnetic field rather than the coupling itself. In contrast, the
energy distribution of an array of localized states cannot be modified
by applying a magnetic field, since the energies of different
localized states have the same magnetic field dependence.

\section{Comparison to the scattering matrix approach}

The effect of dissipation on the persistent current in a one
dimensional ring was examined in previous work already, using the
model depicted in Fig. \ref{fig:compButtikerTunnel}
(b)\cite{Buttiker85:1846}. The analysis of that work was based on the
scattering matrix, that describes the effect of the junction between
the one-dimensional wire and the ideal ring. This approach requires
the use of a continuous basis of wavefunctions in the ring and the
amplitudes inside the ring are related by the Aharonov-Bohm phase
matching condition.  In the frame of the scattering approach only
energetically degenerate wavefunctions inside the ring are mixed by
the coupling. In contrast, the tunnel Hamiltonian, uses the discrete
eigenstates of the isolated ring, that already satisfy the phase
matching, and the coupling leads to a mixing of states with different
unperturbed eigenenergies.

Recently it was shown that calculations based on the scattering matrix
approach or the tunnel Hamiltonian give the same transmission through
an Aharonov-Bohm interferometer with a single-level quantum dot in at
least one of the arms \cite{Kubala02:245301, Kubala02:0212536}.  In
the following we show that for a coupled ring however there are
differences between both approaches. In particular, the scattering
matrix approach used in Ref. \cite{Buttiker85:1846} fails to predict
the formation of additional quasistates at very strong coupling due to
the mixing of eigenstates of the ring that are energetically far from
each other.

The scattering matrix used in Ref. \cite{Buttiker85:1846} depends on a
single free parameter called $\e$ that can be identified with the
coupling strength between ring and reservoir. Furthermore, the authors
limited the coupling strength $0 \le \e \le 0.5$ to keep the matrix
real. The restriction to a real matrix with a single free parameter is
not sufficient to describe a general unitary 3x3 matrix like it was
already anticipated by the authors.  As a consequence of the
restricted range of coupling, the scattering matrix approach used in
Ref. \cite{Buttiker85:1846} fails to describe the mixing of
eigenstates of the ring that are energetically far from each other.
Therefore, even for maximum coupling only one group of quasistates
develops within the scattering matrix approach, with eigenenergies
either at $E=(2n)^2 E_0$ or at $E=(2n+1)^2 E_0$ depending on the
magnetic flux through the ring.

\begin{figure}
\begin{center}
\includegraphics[angle=0,scale=.9]{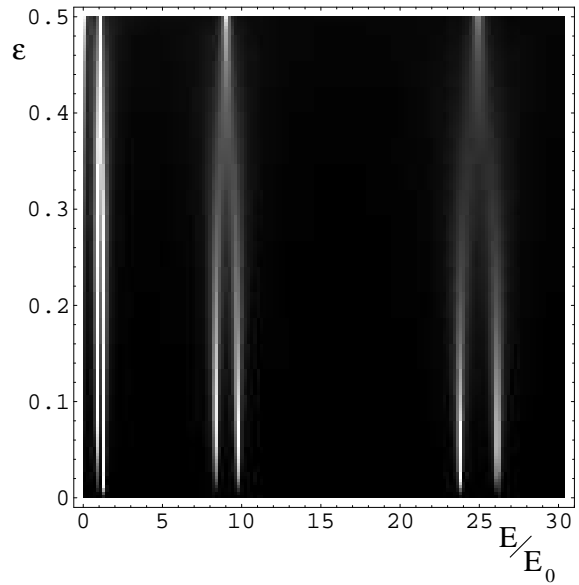}
\end{center}
\caption{Density plot of DOS as a function of energy and coupling
  for the scattering matrix formalism and for a fixed magnetic flux
  $\Phi=0.44 \Phi_0$ (the magnitude of the DOS increases going from
  black to white). For small coupling, the DOS shows maxima at the
  eigenenergies of the isolated ring states, that are grouped in
  pairs. With increasing coupling, the states of each pair approach
  each other and develop new long living states at $E=(2n+1)^2\,E_0$.
  Thereby each long living state is formed at the same coupling
  strength.}
\label{fig:ANrhoEepsilonMultiple}
\end{figure}

Within the accessible range of the coupling strength for the
scattering matrix approach, the results for the DOS and the current
density in the ring are qualitatively the same as the ones obtained in
this paper. In particular, both formalisms show Lorentz broadened
maxima in the density of states around the eigenenergies of the
isolated ring states in the weak coupling limit and the ratio between
current density and DOS is independent of coupling. Furthermore both
formalisms show level attraction as shown in Fig.
\ref{fig:ANrhoEepsilonMultiple} for the scattering matrix approach.
However, within the scattering matrix approach the quasistates develop
at the same coupling strength whereas for the tunnel Hamiltonian the
quasistates with lower eigenenergies are developed at smaller coupling
strength.  This difference between the approaches can be compensated
by choosing an energy dependent coupling strength
$\kappa(E)=\kappa_0\,\sqrt{E}$ (with zero coupling for negative
energies in the reservoir) for which both models nearly coincide.
Another consequence of the energy dependent coupling is that for small
coupling the width of the Lorentz broadened maxima increases like
$\sqrt{E}$.

\begin{figure}
\begin{center}
\includegraphics[angle=0,scale=0.9]{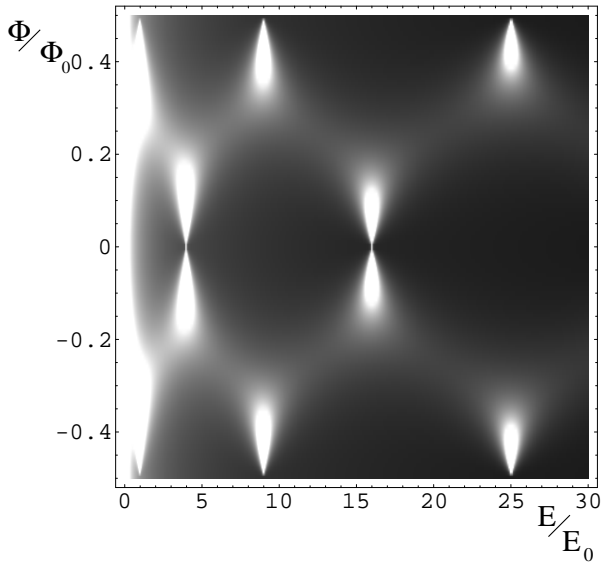}
\end{center}
\caption{Density plot of DOS as a function of energy and magnetic flux
  within the scattering matrix formalism for maximum coupling
  $\e=0.5$. Pronounced maxima are developed at the energies $E=n^2
  E_0$ with n odd or even depending on the flux. This effect gets
  stronger for a flux close to $\Phi=0$ and $\pm\frac{\Phi_0}{2}$,
  corresponding to degenerate eigenenergies. At those values of flux,
  the width of the maximum goes to zero. These results can be
  reproduced within the tunnel Hamiltonian for: $\kappa=0.6\,
  \sqrt{E_0\,E}$}
\label{fig:ANrhoEphi_buttiker}
\end{figure}

\begin{figure}
\begin{center}
\includegraphics[angle=0,scale=.9]{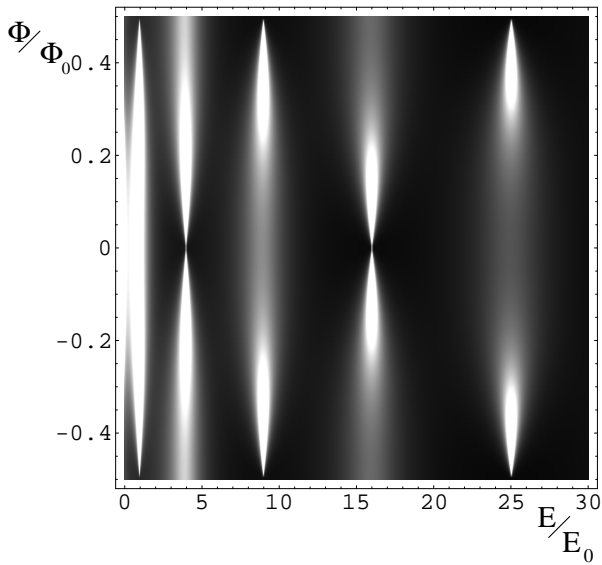}
\end{center}
\caption{Density plot of DOS as a function of energy and magnetic flux
  within the tunnel Hamiltonian formalism. The system is in the strong
  coupling regime and the coupling is energy dependent $\kappa=2\,
  \sqrt{E_0\,E}$. The DOS shows maxima at $E=n^2 E_0$ that are
  particularly pronounced at $\Phi=0$ and $\Phi=\pm\frac{\Phi_0}{2}$,
  where the width of the maximum goes to zero. }
\label{fig:ANrhoEphi_tunnel}
\end{figure}

Furthermore, in both formalisms the coupling is assumed to be
independent of the magnetic flux, while the energy spectrum for the
isolated ring is of course flux dependent. Therefore, the transition
between weak and strong coupling is also flux dependent. This can be
seen in Fig. \ref{fig:ANrhoEphi_buttiker} for the scattering matrix
approach and in Fig. \ref{fig:ANrhoEphi_tunnel} for the tunnel
Hamiltonian approach. The quasistates are the more pronounced the
closer the flux is to the values $\Phi=n \Phi_0$ or $\Phi=\pm
\frac{\Phi_0}{2}$ corresponding to a degenerate energy spectrum.  Fig.
\ref{fig:ANrhoEphi_buttiker} also illustrates that even for maximum
coupling only one group of quasistates is formed within the scattering
matrix description, namely at $E=(2n+1)^2\,E_0$ for
$-0.25<\frac{\Phi}{\Phi_0}<0.25$ or at $E=(2n)^2\,E_0$ for
$0.25<|\frac{\Phi}{\Phi_0}|<0.5$.

\section{Conclusion}
In this paper we studied the energy spectrum and the persistent
current of a ring coupled to a reservoir. Both of these quantities are
accessible in experiments.  The DOS can be measured by means of
optical spectroscopy or by measuring the charging energy
\cite{Lorke00:2223, Warburton00:926}, whereas the current density is
accessible by measuring the magnetization as a function of the Fermi
energy (at low temperatures $j(\mu)\approx \frac{\partial I}{\partial
  \mu}$).  We have shown that for strong coupling the system has a
new, well defined level structure formed by quasistates with sharp
eigenenergies.  The physical mechanism leading to the development of
collective quasistates has been explained in previous work on resonant
tunneling and scattering in solid states \cite{Shahbazyan98:6642} and
is related to the Dicke effect \cite{Dicke53:472, Dicke54:99}.  We
derived analytical formulas for the DOS and the current density in the
ring, and analyzed their dependence on the tunneling strength between
ring and reservoir as well as on the number of coupled states.
Thereby the number of coupled states depends on the geometrical form
of the contact.  It was shown that for a finite number of coupling
states, the persistent current is rather robust against coupling and
does only vanish if all ring states couple to the reservoir.  Our
model can reproduce the results obtained in previous work based on the
scattering matrix approach and furthermore it describes how the system
reaches a new stable energy spectrum in the limit of strong coupling.

Finally we note, that the experimental realization of strong tunnel
coupling is achieved by creation of a quantum well in the contact area
between the reservoir and the ring. The quantum well modifies the
dynamical motion of the charge carriers, which manifests itself in the
formation of sharp quasistates is the ring.

\section{Appendix}
For the calculation of the DOS and the current density in the ring it
is useful to express $\xi(E)=\sum_{m=m_1}^{m_2} \frac{1}{E-\e_m}$ with
the help of the digamma function
$\psi(z)=\frac{\Gamma\,'(z)}{\Gamma(z)}$.

\begin{eqnarray}
\xi(E)&=&
\frac{\psi^-_{m_1}(E)-\psi^+_{m_1}(E)-\psi^-_{1+m_2}(E)+\psi^+_{1+m_2}(E)}{4\sqrt{E_0 E}}\notag\\
\end{eqnarray}
with 
\begin{eqnarray}
\psi^-_{m}(E)&=&\psi\left(m-\sqrt{\frac{E}{4 E_0}}+ \frac{\Phi}{\Phi_0}\right),\\ 
\psi^+_{m}(E)&=&\psi\left(m+\sqrt{\frac{E}{4 E_0}}+ \frac{\Phi}{\Phi_0}\right).
\end{eqnarray}

With the help of {Eq. (\ref{rhogeneral}), (\ref{jgeneral})} the DOS
and the current density in the ring can be calculated, using the
polygammafunction $\psi^{(1)}(z)=\frac{\partial}{\partial z} \psi(z)$
.
\begin{eqnarray}
\rho(E)&=&-\kappa\left(-2 \sqrt{E_0}\xi(E)
+\sqrt{E}\left(-\psi^{-\,(1)}_{m_1}(E)+\right.\right.\notag\\
&+&\left.\left. \psi^{-\,(1)}_{1+m_2}(E)
- \psi^{+\,(1)}_{m_1}(E)
+ \psi^{+\,(1)}_{1+m_2}(E)\right)\right)/\notag\\
& &\left(16 E_0 \pi E^{\frac{3}{2}}\left(1+\kappa^2 \xi(E)^2\right)\right),\\
\notag\\
j(E)&=&
- \frac{ \kappa}{\Phi_0}\left(\psi^{-\,(1)}_{m_1}(E)
-\psi^{-\,(1)}_{1+m_2}(E)- \psi^{+\,(1)}_{m_1}(E)
+ \right.\notag\\
&+&\left.\psi^{+\,(1)}_{1+m_2}(E)\right)/\left(4 \pi \sqrt{E_0 E}\left(1+\kappa^2 \xi(E)^2\right)\right)\notag\\
\end{eqnarray}
with 
\begin{eqnarray}
\psi^{-\,(1)}_{m}(E)&=&\psi^{(1)}\left(m-\sqrt{\frac{E}{4 E_0}}+ \frac{\Phi}{\Phi_0}\right),\\ 
\psi^{+\,(2)}_{m}(E)&=&\psi^{(1)}\left(m+\sqrt{\frac{E}{4 E_0}}+ \frac{\Phi}{\Phi_0}\right)\quad.
\end{eqnarray}

\section{Acknowledgments}

The authors are grateful to D. Pfannkuche for illuminating
discussions.  Financial support form SFB 508 is gratefully
acknowledged.

\bibliographystyle{apsrev}
\bibliography{cringref}

\end{document}